# Local and global dynamics in polypropylene glycol / silica composites

R. Casalini and C.M. Roland

Naval Research Laboratory, Chemistry Division, Code 6120, Washington DC 20375-5342

**Abstract**

The local segmental and global dynamics of a series of polypropylene glycol / silica nanocomposites were studied using rheometry and mechanical and dielectric spectroscopies. The particles cause substantial changes in the rheology, including higher viscosities that become non-Newtonian and the appearance of stress overshoots in the transient shear viscosity. However, no change was observed in the mean relaxation times for either the segmental or normal mode dynamics measured dielectrically. This absence of an effect of the particles is due to masking of the interfacial response by polymer chains remote from the particles. When the unattached polymer was extracted to isolate the interfacial material, very large reductions in the relaxation times were measured. This speeding up of the dynamics is due in part to the reduced density at the interface, presumably a consequence of poorer packing of tethered chains. In addition, binding of the ether oxygens of the polypropylene glycol chains truncates the normal mode, which shifts the corresponding relaxation peak to higher frequencies.

**Introduction**

Over the past decade there has been a plethora of activity directed to studying and exploiting polymer nanocomposites [1,2,3,4,5,6]. The mechanical and other properties of polymers can be greatly affected by the presence of small particles. In addition to redistributing strains locally (hydrodynamic interaction), particulate fillers can perturb the segmental dynamics, as manifested by changes in the temperature and intensity of the glass transition. Very generally, the expectation is that the polymer material comprising the interface with the particles will experience effects analogous to those observed in thin films, with the potential for additional effects due to specific interfacial interactions [7,8,9,10].

Less work has been reported concerning the effect of nanoparticles on the global dynamics. The rheology of nanocomposites is of obvious import, given the enormous specific surface area of nanoparticles [11,12]. Frank et al. [13] observed substantial decreases in the lateral diffusion coefficient of polymer chains in thin films. On the other hand, Kremer et al. [14] observed no change in the normal mode frequency for polyisoprene confined to thin films, but did report the appearance of a new mode, ascribed to motion of subchains associated with chain segments confined at the interface. For nanocomposites of polymer blends, the viscosity in the vicinity of nanoparticles was reported to be lower than for the bulk material [15], ascribed to alteration of the composition of the blend near the nanoparticle surface. It also appears that for high molecular weight polymers, thin film confinement may cause a reduction in entanglements, with consequent decrease in viscosity and related rheological properties [16,17,18]. However, increased entanglement has also been observed in simulations [19], and experimentally a reduced compliance was found for thin films of polystyrene [20]. The related problem of diffusion of polymer chains through nanopores has been studied theoretically and numerically by various groups [21,22,23,24,25].

Generally, bulk measurements do not show large changes in $T_g$ or the segmental dynamics, especially when the interfacial material represents only a fraction of the total polymer [26,27]. The details of the particle-polymer interaction, as well as any densification of the interfacial material, appear to be the dominant mechanisms for perturbation of the local dynamics. However, it is an experimental challenge to decouple the effect of the interface on the dynamics from that of the bulk (unbound) material. The situation is similar to miscible blends, in which the components exert reciprocal influences. Simulations can overcome the problem of the behavior of the interfacial material being swamped by that of the bulk, by enabling a focus on only those chains in proximity to the particles [28,29,30,31,32,33]. Simulations suggest that attraction to the particle surface expands the size of the interfacial chains [34], presumably affecting their motion. For the rheological properties, simulations suggest a reduction in entanglements as the origin of enhanced mobility [35].

One of the more common particulates employed in nanocomposites is silica. Various groups have reported at most only small effects on the local segmental dynamics of silica nanocomposites [36,37,41,38], although there can be detectable changes on the lower frequency side of the segmental dispersion [36]. Cheng et al. [39] found increases in the glass transition temperature and fragility in silica-glycerol mixtures, ascribed to densification in the interfacial region. Addition of silica nanoparticles was reported to reduce the intensity of the glass transition of poly-2-vinylpyridine due to restricted segmental mobility of polymer segments at the interface [40]; on the other hand, for silica nanocomposites of polyvinylacetate no change was observed in the magnitude of the property changes at $T_g$, beyond that due to the replacement of polymer by the filler [41]. Atomic Force Microscopy showed restricted mobility for styrene-butadiene copolymer at the interface with silica particles [42]. From quasielastic neutron scattering, Masui et al. [43] concluded that the primary effect of silica nanoparticles was a longer residence time between diffusive jumps.

In this work we investigate the effect of silica nanoparticles on polypropyleneglycol (PPG). This particular polymer was chosen because it has a dipole moment parallel to the chain, which enables the global chain motions to be measured by dielectric spectroscopy. To resolve the dynamics at the interface, we extracted the unbound polymer. We find that after extraction, the local segmental and the normal mode relaxation times are both reduced by the particulate reinforcement (i.e., faster dynamics); interestingly, the changes in the global motions are greater than the changes of the segmental dynamics. We also observe in steady shear experiments an overshoot peak for the nanocomposite. This stress overshoot exhibit a delayed recovery after cessation and restarting the shear flow; the time scale for the full magnitude of the overshoot to be observed is substantially longer than the relaxation times associated with the rheological properties. We ascribe this shear-induced structural change, which requires strains of ~500%, to interactions between the silica particles.

**Experimental**

Polypropylene glycol (PPG) (weight average molecular weight = 4000 Da) was obtained from Polysciences Inc. Because of its hydrophilic nature, care was taken to dry the material (by heating at 80°C *in vacuo*) before measurements, and the latter were done either under vacuum or a dry nitrogen atmosphere. Silica (Si) nanoparticles (diameter ~ 12.5 nm) were obtained from Nissan Chemicals. The particles are functionalized with hydroxyl groups (5 to 8 OH per $nm^2$) and received as an isopropanol solution. To prepare the nanocomposites, the polymer was dissolved in isopropanol and mixed with the Si solution; after sonication, the isopropanol was removed by heating at 80°C in vacuum.

Dynamic mechanical and shear flow measurements employed with an Anton Paar MCR 502 rheometer, using a cone and plate fixture (50 mm diameter with 1° cone angle). Dielectric relaxation spectroscopy was carried out with the sample between cylindrical electrodes (16 mm diameter) and a 0.1 mm. PTFE spacer to maintain a constant thickness. A Novocontrol Alpha analyzer was used at frequencies from $10^{-3}$ to $10^6$ Hz. The temperature was controlled using a closed cycle helium cryostat (Cryo Ind.).

**Results**

**Rheology.** Figure 1 shows the dynamic viscosity for the neat PPG and four nanocomposites. At sufficiently high silica loadings, the behavior becomes non-Newtonian. For the highest concentration silica, there is power law behavior with an exponent = -0.6. This non-linearity is seen more clearly in the inset, showing the viscosity as a function of silica volume fraction, $\phi$. For spherical particles, the hydrodynamic contribution to the viscosity increase is described by the equation of Guth and Gold [44] as modified for solvation effects [45,46]

$$\eta(\phi) = \eta_0 \left[ 1 + 2.5(f\phi) + 14.1(f\phi)^2 \right] \tag{1}$$

where $\eta_0$ is the viscosity absent filler and $f$ accounts the effective "growth" of the particles due to bound or occluded polymer. Fitting eq. (1) to the data for $\phi < 0.13$, we obtain $f = 4.4$, which is in the range reported for carbon black composites in which there is substantial interfacial polymer [1].

The steep rise in viscosity at higher filler content (Fig. 1) reflects the large hydrodynamic effect due to the enormous interfacial area (the viscosity increase for $\phi$=0.14 is at least fifteen-fold larger than obtained with conventional silica [47]), augmented by interparticle interactions. The magnitude of the latter can be assessed from nonlinearity in the mechanical response, for example, a strain-dependent dynamic modulus [48,49]. The relatively low molecular weight of the PPG herein, however, permits another approach based on the response to steady shear flow.

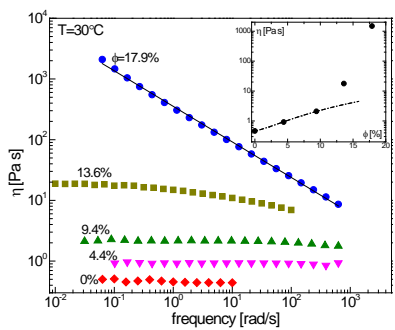

Figure 1. Dynamic viscosity for PPG having the indicated silica concentration. The inset shows $\eta$ (0.1 rad/s) vs. $\phi$. The dashed line is eq. (1) fit to the lower filler levels, yielding $f$=4.4.

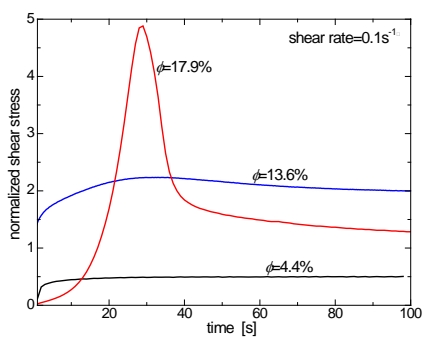

Figure 2. Transient shear stress normalized by the long-time steady state value (curves displaced vertically for clarity). At higher levels of silica, an overshoot becomes apparent, especially prominent for the largest $\phi$. The respective temperatures were chosen to optimize the measurement (T=-50, -20, and 100°C for lowest to highest concentration); the height of the overshoot varies weakly with temperature.

As seen in Fig. 1, at low levels of silica the PPG exhibits Newtonian behavior. This is the general result for low molecular weight, neat polymers. Materials with time-dependent structure, however, can exhibit stress overshoots and subsequent shear thinning, as the structure is lost upon onset of flow. The most common example of this behavior is flow-induced disentanglement of polymer chains, which causes a transient overshoot upon startup of the shearing [50,51,52,53]. Laboratory characterization of this phenomenon is limited to materials with a low concentration of entanglements; otherwise there is melt fracture and non-uniform flow [54,55]. Shown in Figure 2 is the transient stress measured for the PPG nanocomposites during initiation of steady shear flow. At the highest silica level ($\phi$=0.179) the stress passes through a distinct maximum versus shearing time, followed by a steady state response. This overshoot peak is also present, albeit weak, in the PPG with 13.6% silica. These are the two nanocomposites exhibiting shear-thinning behavior in Fig. 1. The stress overshoot is absent for samples with less silica.

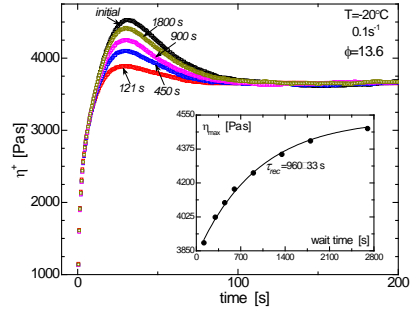

Figure 3. Transient viscosity in the nanocomposite measured at various rest intervals between shearing. The recovery of the overshoot maximum is shown in the inset; the solid line is an exponential fit, yielding the indicated recovery time constant.

If the shear flow is stopped and then immediately resumed, there is no maximum in the transient stress. Nevertheless, the disruption of structure underlying the overshoot is a physical process, and thus reversible. This is illustrated in Figure 3, showing the transient viscosity, $\eta^+$, measured after various rest intervals prior to resumption of the shearing. A time-constant, $\tau$, can be obtained for the growth of the overshoot, by describing the peak magnitude, $\eta^+_{max}$, measured for various rest times with an exponential function

$$\eta^+_{max}(t) = a \exp\left(\frac{-t}{\tau}\right) + \eta^+_{max,\infty} \qquad (2)$$

In eq.(2) $\eta^+_{max,\infty}$ is the maximum of the transient viscosity after an extended recovery period (exceeding one hour near ambient temperature for the highest silica concentration studied); $\eta^+_{max,\infty}$ thus equals the value of the initial startup shear flow measurement. The fit of eq. (2) to data for $\phi$=0.136 silica is shown in the inset to Fig. 3, yielding $\tau$=960 s at -20°C. (The temperatures chosen for the experiments reflected the need to have a sufficiently large viscosity and kinetics slow enough for facile measurement.) The steady state viscosity for this sample at -20°C was 3.68 kPa s. Estimating a time constant from the Maxwell relation

$$\tau_M = \eta / G_\infty \qquad (3)$$

in which $G_\infty$ is the high frequency limiting value of the modulus, we obtain a value ($\tau_M < 10^{-5}$ s) several orders of magnitude smaller than the recovery time. The implication is that recovery of the overshoot in the transient viscosity cannot be governed by the polymer chain dynamics. To identify the mechanism for the overshoot, in Figure 4 are Arrhenius plots of the recovery time constant and for the viscosity. The activation energy for the latter, 83.5 ± 1.6 kJ/mol, is significantly larger than for the overshoot recovery, 39.3 ± 2.2 kJ/mol. This disconnect between the temperature dependences of the two processes is consistent with the idea that that the recovery is not controlled by the chain dynamics. The inference is that particle-particle interactions give rise to a structure that elevates the dynamic viscosity measured at low strain amplitudes (Fig. 1); however, the connectivity is fragile, its labile nature yielding an overshoot in the transient viscosity.

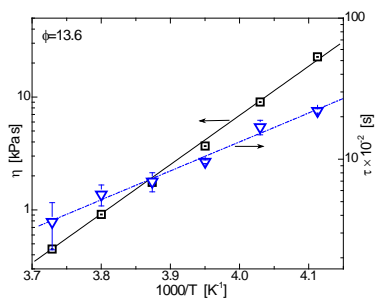

Figure 4. Arrhenius plots for the nanocomposite having the indicated silica content: (left) viscosity (squares); (right) overshoot recovery time constant (inverted triangles). The latter has a smaller activation energy.

**Dielectric relaxation.** Since PPG has a dipole moment parallel to the chain, we can further probe the chain dynamics, as well as the local segmental relaxation, using dielectric spectroscopy. In Figure 5 are shown representative dielectric relaxation spectra for the neat PPG and the nanocomposite with $\phi$=0.136; the measurements are at nearly the same temperature. Two peaks are observed, corresponding to the local segmental and global chain modes, at higher and lower frequencies respectively. Fitting the permittivity spectra to the Kohlrausch function [56]

$$\varepsilon^*\left(f\right) = \Delta\varepsilon L\left[-\frac{d\left(\exp\left[-(t/\tau_K)^\beta\right]\right)}{dt}\right] \tag{4}$$

where $f$ is the frequency, $\tau_K$ a mean relaxation time, $\Delta\varepsilon$ the dielectric strength, $\beta$ the shape parameter, and $L$ indicates the Laplace transform. We simultaneously fit both the real and imaginary part of $\varepsilon^*$ for both dispersions, including a term for the dc-conductivity, $\sigma_{dc}$ ($\propto f^{-1}$). The normal mode peak is broader for the nanocomposite ($\beta$=0.51) compared to that for neat PPG ($\beta$=0.68), whereas the segmental dispersion ($\beta$=0.50±0.01) is unaffected by the silica. The obtained relaxation times are shown in Figure 6. Clearly there is no observable effect on the average relaxation dynamics due to the present of the silica particles.

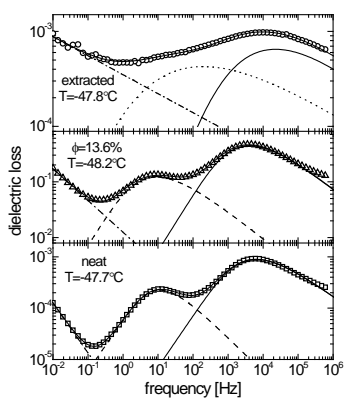

Figure 5. Dielectric spectra of (bottom to top): neat PPG; nanocomposite; nanocomposite after extraction of unattached polymer. The lines represent fits of the Kohlrausch function to each of the two dispersions, along with a power-law at low frequencies due to dc-conductivity.

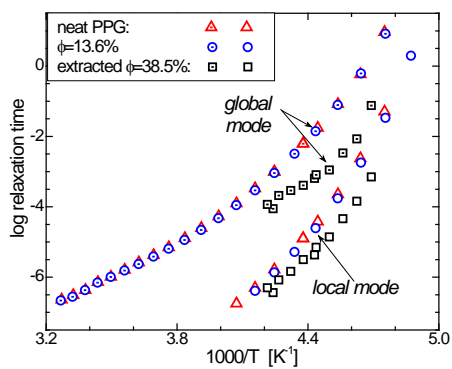

Figure 6. Local segmental (dotted symbols) and dielectric normal mode (open symbols) relaxation times for the neat PPG (triangles), a nanocomposite (circles), and the nanocomposite after extraction of the unbound polymer (squares).

To emphasize the contribution of the interfacial polymer to the dynamic behavior, we extracted the nanocomposite with hexane. This removed all soluble PPG; the remaining polymer (18.2% by weight) is attached to the silica via the surface hydroxyl groups. The dielectric spectrum of this extracted material, containing only silica and bound PPG, is also shown in Fig. 5. Although the measurements are noisier (the sample has no mechanical integrity), it is evident that both the segmental relaxation and the normal mode peaks shift to higher frequency. The speeding up of the dynamics is greater at lower temperatures, with the normal mode relaxation time reduced by as much as a factor of fifty. The effect on the segmental mode is smaller but still substantial, about a fivefold reduction in the relaxation time after extraction. This acceleration of the motions corresponds to temperature increases of about 10K for the normal mode and 4K for the segmental mode. Notwithstanding the greater mobilities, both the local and global relaxation dispersions are much broader for the extracted sample, $\beta$=0.29 and 0.24 respectively.

**Discussion and Summary**

Nanoparticles can affect both the local segmental and the global dynamics of polymers. Herein silica particles were found to exert a very large effect on the viscosity of PPG and give rise to a prominent overshoot in the transient viscosity during startup of steady shear flow. The retarded recovery of the startup transient indicates it involves interparticle interactions that are largely decoupled from the local viscosity. Notwithstanding these effect of the silica on the rheology of the PPG, the dielectric normal mode measured for the neat PPG and the nanocomposite were equivalent. However, when the unbound polymer was removed by solvent extraction, the global dynamics of the residual PPG, which is adhered to the particles, was found to be more than an order of magnitude faster than for the bulk polymer. A similar result was found for the segmental relaxation, although there was less speeding up of the more local dynamics. To the extent that the local friction factor for the segmental dynamics can be identified with the friction factor governing the global dynamics, as assumed in classical models for the chain dynamics [57], changes in the two processes should be comparable. However, truncation of the normal modes may underlie the larger effect of the particles on the global motions. If the etheric oxygens in the chain are connected to the silica hydroxyl groups via hydrogen bonds, the normal mode would involve smaller sections of the polymer chain, and thus be faster. This would also account for the broader distribution of relaxation times seen for the normal mode. However, if this truncation is the cause of the greater shift of the normal mode peak compared to the segmental relaxation, the effect artifactual rather than actual greater enhancement of the global mobility.

Since the segmental dynamics of the bound chains are also faster than for the neat polymer, another mechanism is operative. The usual consequence of specific interactions with filler particles, to constrain and retard motions, must be absent herein, or at least overwhelmed by other effects. We can assess the density in the vicinity of the particles by comparing the mass density of the nanocomposite to the prediction assuming additivity of the component volumes. From the densities of the components, $\rho$ = 1.004 and 2.2 g/ml for PPG and silica respectively, we calculate 1.218 g/ml for $\phi$ = 0.179. We measured, however, 1.12$\pm$0.02 g/ml for this nanocomposite sample; that is, the interface has a significantly lower density than the bulk. This result leads to the conclusion that, notwithstanding their being tethered to particles, the interfacial chains experience reduced crowding, presumably the result of poor packing. This would account for faster segmental dynamics.

**Acknowledgments**

This work was supported by the Office of Naval Research.

--- for Table of Contents use only ---

Local and global dynamics in polypropylene glycol / silica composites

R. Casalini and C.M. Roland

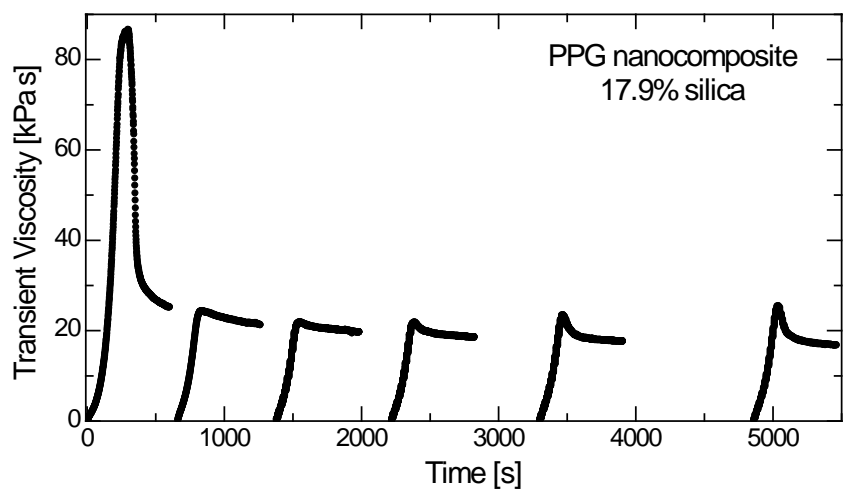